\documentclass{elsart}

\begin{document}

\begin{frontmatter}


\title{Two-time scale subordination in physical processes with long-term memory}

\author[label1]{Aleksander Stanislavsky\corauthref{cor1}}
\author[label2]{, Karina Weron}
\corauth[cor1]{\it{E-mail address}: \rm{alexstan@ri.kharkov.ua}}
\address[label1]{Institute of Radio Astronomy, Ukrainian National Academy of
Sciences,\\ 4 Chervonopraporna St., 61002 Kharkov, Ukraine}
\address[label2]{Institute of Physics,  Wroc{\l}aw University of Technology,\\ Wyb.
Wyspia$\acute{n}$kiego 27, 50-370 Wroc{\l}aw, Poland}

\begin{abstract}
We use the two-time scale subordination in order to describe
dynamical processes in continuous media with a long-term memory.
Our consideration touches two physical examples in detail. First
we study a temporal evolution of the species concentration for the
trapping reaction in which a diffusing reactant is surrounded by a
sea of randomly moving traps. The analysis is based on the
random-variable formalism of anomalous diffusive processes. We
find that the empirical trapping-reaction law, according to which
the reactant concentration decreases in time as a product of an
exponential and a stretched exponential function, can be explained
by the two-time scale subordination of random processes. Another
example is connected with a state equation for continuous media
with memory. If the pressure and the density of a medium are
subordinated in two different random processes, then the ordinary
state equation becomes fractional with two time scales. This
allows one to arrive at the state equation of Bagley-Torvik type.

\end{abstract}

\begin{keyword}
Subordination \sep Nonexponential relaxation \sep Trapping
reaction \sep Anomalous diffusion \sep Continuous medium with
memory

\PACS 02.50.-r \sep 82.40.-g \sep  02.50.Ey

\end{keyword}

\end{frontmatter}

\section{Introduction}
The problem of trapping reactions has a long and active history
tracked in the literature (see, for example, \cite{TW,BL,AH}). The
reaction dynamics has been also studied over the past decade
\cite{SBSL,MOBC,BMB,YL1,YL2}. In the traditional version a
reactant ($A$) walks in a medium randomly doped with static traps
($B$) so that when they meet, the reactant disappears ($A+B\to
B$). This has served as a starting point for the formulation of
several models describing the behavior of more complex systems. An
important variation of the basic trapping problem is that the
traps become diffusive too. When both species ($A$ and $B$) are
(sub)diffusive, the temporal evolution of the system looks like a
random motion (of walker $A$) in a random sea of traps (trapping
random walkers $B$). One approach for the analysis of such
processes is based on the continuous time random walk (CTRW) in
which space random jumps follow among subsequent random waiting
times \cite{KBZ,MK}. Another approach is stated on the fractional
diffusion equation, which describes the probability density of
finding the particle at position $x$ at time $t$ \cite{SW,IP}.
Recently, the relationship between the CTRW, stable distributions
and the fractional calculus has been established exactly
\cite{MS,S0}. It turns out that if the CTRW is represented as a
subordination of a space random (Brownian or L$\acute{\rm e}$vy)
process by a inverse-time $\alpha$-stable L$\acute{\rm e}$vy
process, the probability density of a walker position is governed
by a fractional Fokker-Planck equation. The approach was also
extended on anomalous relaxation processes \cite{Sok,S1,MW1,MW2}.
The aim of the given paper is to present a development of the
subordination formalism to the complex relaxation processes. We
are going to consider a time evolution of the concentration in the
trapping reactions when the relaxation function takes the form of
a product of an exponential and a stretched exponential function.
The experiment supporting this empirical law is described in the
paper of Djordjevi$\check{\rm c}$ \cite{D}. The author studied the
problem of an electron transfer from the methyl viologen radical
cations to the colloidal platinum particle in a water solution. In
that reaction the monocations MV$^+$ became the dications
MV$^{2+}$ and the dependence of MV$^+$ concentration on time
appeared to follow the combined exponential-stretched-exponential
decay law. In particular the experiment has stated the relaxation
curve following $\log C(t)=-0.0011825t - 0.0608t^{0.6}$ such that
the cautious estimation of the exponent of the second term is
$0.56\pm 0.06$. Next, Djordjevi$\check{\rm c}$ also noticed that
the contribution of the second term is about $90 ^o\!/\!_o$ of
$\log C(t)$. One of the aims of his paper was to verify
experimentally the fact of factorization. The task was realized
with success. We believe that this experiment is good, and the
data fitting gives the physical dependence clearly.

The paper is organized as follows. In Sec.~\ref{par2} we present
important features of subordinate random processes with different
subordinators. They are directly connected with the anomalous
diffusion. In the dependence of the subordinator form the
corresponding subordinate processes have a different evolution in
time. In our analysis we use the peculiarity of relaxation
functions. From the paper \cite{MW2} it is known what subordinator
leads to the stretched exponential law. In Sec.~\ref{par3} we show
that for the relaxation law to take the combined form, the
subordinator should contain two internal-time scales. One of them
gives the exponential relaxation, and another contributes the
stretched exponential function. In the spirit of the
consideration, Section~\ref{par4} is devoted to the study of
continuous media with a long-term memory. We consider such media
that are characterized by the two time scales in the state
equation. Section~\ref{par5} presents a summary of results.

\section{Stretched exponential response and its subordinator}\label{par2}
If a complex physical system consisting of identical objects
(dipoles, charges and so on) undergoes an irreversible transition,
say, from state $A$ to state $B$ at random instances of time, then
the transition can be characterized by the probability in the form
\begin{displaymath}
{\rm Pr}(\theta\geq t)=\exp\left(-\int^t_0 r(y)\,dy\right),
\end{displaymath}
where the non-negative quantity $r(y)$ denotes the system's
transition rate, i.e., the transition probability intensity for
transition of the system as a whole. This formula simply follows
from a two-stage master equation (see, for example, \cite{Kamp}).
The probability ${\rm Pr}(\theta\geq t)$ shows that a considered
object will remain in state $A$ until time $t$ . In general, the
transition rate $r(y)$ is time-dependent because of random impacts
affecting each object. If one knows the explicit form of $r(y)$,
the value ${\rm Pr}(\theta\geq t)$ can be derived. In the simplest
case, when the quantity $r(y)=b_0={\rm const}$ is
time-independent, the above formula recovers the classical
exponential relaxation law
\begin{displaymath}
\phi(t)={\rm Pr}(\theta\geq t)=e^{-\omega_pt},
\end{displaymath}
where $\omega_p=b_0$ is a characteristic material constant. The exponential law, however
hardly ever found in nature, is widely accepted for description of various relaxation
data. Such a model of relaxation assumes that the relaxation rate (inverse of the
relaxation time) of each object is constant. Although it may be different from one object
to others, the mean of the effective relaxation rate (representing the inherent
stochastic nature of the relaxation process) has to take a finite value. Nevertheless,
the extensive experimental investigations in a wide frequency-time domain have shown
\cite{J1,J2} relatively large deviations from the exponential relaxation law. It has been
found that for many materials the relaxation response follows the stretched exponential
pattern
\begin{displaymath}
\phi(t)=e^{-(\omega_pt)^\alpha},\qquad 0<\alpha<1.
\end{displaymath}
In this case, on contrary, the system's transition rate is essentially time dependent,
and the mean of the effective relaxation rate becomes infinite \cite{JW}.

The time dependence of the system's transition rate as well as the statistical properties
of the effective relaxation rate clearly depend on the defect-diffusion relaxation
mechanism in the system under consideration \cite{MK}. Hence, the exponential and the
nonexponential relaxation can be modeled by means of the diffusive behavior of the
systems as a whole. Of course, the exponential relaxation and nonexponential one
correspond to different diffusion processes.

Consider a sequence ${T_i}$, $i=1,2,\dots$ of non-negative, independent, identically
distributed random variables which represent waiting-time intervals  between subsequent
jumps of a particle. The random time interval of $n$ jumps in space is written as
\begin{equation}
T(n)=\sum_{i=1}^n T_i,\qquad T(0)=0.\label{eq11}
\end{equation}
The number of the particle jumps performed up to time $t>0$ is determined by the largest
index $n$ for which the sum of $n$ interjump time intervals does not exceed the
observation time $t$
\begin{displaymath}
N_t=\max\{n:T(n)\leq t\}.
\end{displaymath}
The process $N_t$ is called the renewal process, or else, the counting process. The
position of the particle (i.e., the total distance) after $N_t$ jumps becomes
\begin{equation}
R(N_t)=\sum_{i=1}^{N_t}R_i,\label{eq11a}
\end{equation}
where $R_i$ are independent, identically distributed variables giving both the length and
the direction of the {\it i}-th jump. The process (\ref{eq11a}) is just known as the
CTRW.

Assume that the time intervals $T_i$ belong to the domain of attraction of a completely
asymmetric stable distribution with the index $0<\alpha<1$. The generalization of the
central limit theorem yields the continuous limit of the random sum (\ref{eq11}), namely
\begin{displaymath}
a^{-1/\alpha}T([a\tau])\stackrel{d}{\rightarrow}U(\tau)\qquad {\rm as}\qquad a\to\infty,
\end{displaymath}
where $U(\tau)$ is a strictly increasing $\alpha$-stable L$\acute{\rm e}$vy process,
$a>0$ parameter, $[x]$ denotes the integer part of $x$ and
``$\stackrel{d}{\rightarrow}$'' means ``tends in distribution''. Similarly, let the jumps
$R_i$ belong to the domain of attraction of a $\gamma$-stable distribution
$S_{\gamma,\beta}(x), 0<\gamma\leq 2, |\beta|\leq 1$ so that
\begin{displaymath}
a^{-1/\gamma}R([a\tau])\stackrel{d}{\rightarrow}X(\tau)\qquad {\rm as}\qquad a\to\infty,
\end{displaymath}
where $X(\tau)$ is a $\gamma$-stable L$\acute{\rm e}$vy process. If $\gamma=2$, the
latter process is the classical Brownian motion. Both the process $U(\tau)$ and the
process $X(\tau)$ are indexed by the internal time $\tau$. The time is not the real,
physical time. In order to find a particle position at the observable time $t$, we have
to introduce the notion of the inverse-time $\alpha$-stable L$\acute{\rm e}$vy
subordinator $V_t$ relating the internal $\tau$ and the observable $t$ times
\begin{equation}
a^{-\alpha}N_{at}\stackrel{d}{\rightarrow}V_t=\inf\{\tau:U(\tau)>t\}\qquad
{\rm as}\qquad a\to\infty.\label{eq12}
\end{equation}
Then the continuous limit of the CTRW process obtains the following form
\begin{equation}
a^{-\alpha/\gamma}R(N_{at})\approx(a^\alpha)^{-1/\gamma}R([a^\alpha
V_t])\stackrel{d}{\rightarrow}X(V_t) \qquad {\rm as}\qquad
a\to\infty\label{eq13}
\end{equation}
known as the anomalous diffusion process. In other words, the scaling limit of the CTWR
leads to the anomalous diffusion process $X(V_t)$ in which the parent process $X(\tau)$,
replacing the random discrete-time jumps $R(n)$, is subordinated by the directing process
$V_t$, replacing the counting process $N_t$. It has been rigorously proved that the
probability density of $X(V_t)$ is the solution of well-known fractional diffusion
equation \cite{MS,S0}. As the processes $X(\tau)$ and $V_t$ are independent, following
the total probability formula, the probability density $p_\alpha(x,t)$ of $X(V_t)$ can be
written as
\begin{displaymath}
p_\alpha(x,t)=\int_0^\infty f_\alpha(\tau,t)\,g(x,\tau)\,d\tau\,,
\end{displaymath}
where $f_\alpha(\tau,t)$ and $g(x,\tau)$ are the probability density of $V_t$ and
$X(\tau)$ respectively. Similarly, the Fourier transform $\tilde
p_\alpha(k,t)=\left\langle\exp(ikX(V_t))\right\rangle$ and the Laplace transform $\bar
p_\alpha(k,t)=\left\langle\exp(-kX(V_t))\right\rangle$ are given by
\begin{eqnarray}
\tilde p_\alpha(k,t)=\int_0^\infty f_\alpha(\tau,t)\,\tilde
g(k,\tau)\,d\tau\,,\label{eq14a}\\ \bar
p_\alpha(k,t)=\int_0^\infty f_\alpha(\tau,t)\,\bar
g(k,\tau)\,d\tau.\label{eq14b}
\end{eqnarray}
Here $k>0$ has the physical meaning of a wave number.

The analysis of the properties of the diffusion front (i.e., the asymptotic distribution
of the particle position in time $t$) allows one to determine \cite{S1,MW1} the
one-parameter Mittag-Leffler relaxation function (corresponding to the Cole-Cole law in
the frequency domain), namely
\begin{equation}
\phi(t)=\left\langle
e^{-kX(V_t)}\right\rangle=E_\alpha(-c_{\alpha,\gamma}k^\gamma
t^\alpha),\label{eq15}
\end{equation}
where $E_\alpha(x)=\sum_{k=0}^\infty x^n/\Gamma(n\alpha+1)$ is the Mittag-Leffler
function, $c_{\alpha,\gamma}$ is constant
($\omega_p=(c_{\alpha,\gamma}k^\gamma)^{1/\alpha}$). For the relaxation function to take
another form, the subordinator of the process $X(t)$ should be changed.

With that end in view the papers \cite{Sok,MW2} show that if the subordinator is a fully
asymmetric L$\acute{\rm e}$vy $\alpha$-stable process with the self-similar property
\begin{equation}
V^{(\alpha)}_{bt}\stackrel{d}{=}(bt)^{1/\alpha}V_1^{(\alpha)},\label{eq16}
\end{equation}
where ``$\stackrel{d}{=}$'' reads ``equal in law'', the type of relaxation remains simple
exponential
\begin{equation}
\left\langle e^{-kV^{(\alpha)}_t}\right\rangle=e^{-c_\alpha
k^\alpha t},\qquad 0<\alpha<1.\label{eq17}
\end{equation}
This means that there are different scenarios leading to the
exponential term in the relaxation function. It can be the
Brownian as well as a deterministic process.

Nevertheless, if a new subordinator $\overline{V}^{(\alpha)}_t$ is defined as \cite{MW2}
\begin{equation}
\overline{V}^{(\alpha)}_{t}:\stackrel{d}{=}ct
V_1^{(\alpha)},\qquad c>0,\label{eq18}
\end{equation}
then the relaxation law decays otherwise. This process is
constructed on the $\alpha$ stable L$\acute{\rm e}$vy process
$V_t^{(\alpha)}$ multiplying both sides of its self-similarity
property
\begin{equation}
{V}^{(\alpha)}_{ct}\stackrel{d}{=}(ct)^{1/\alpha}
{V}_1^{(\alpha)},\qquad c>0,\label{eq19}
\end{equation}
by $(ct)^{1 - 1/{\alpha}}$. Both processes $V^{(\alpha)}_t$ and
$\overline{V}^{(\alpha)}_t$ have different properties. The old one
$V_t^{(\alpha)}$ is an $1/{\alpha}$ self-similar L$\acute{\rm
e}$vy process, and this means that it is an $\alpha$ stable
L$\acute{\rm e}$vy process. While the new one is an $\alpha$
stable process, not Levy (i.\ e.\ it does not have independent
increments). Note that any L$\acute{\rm e}$vy process is a process
with independent and stationary increments, whereas the $\alpha$
stable is a process for which all finite dimensional distributions
are stable. Hence, the $\alpha$ stable L$\acute{\rm e}$vy process
$V_t^{(\alpha)}$ has independent and stationary increments as well
as all finite dimensional distributions stable. It is a broad
class of random processes (see Table in \cite{MW2}).

It would be useful to notice that the self-similarity of the process $V_t^{(\alpha)}$ can
be also written as
\begin{displaymath}
{V}^{(\alpha)}_{at}\stackrel{d}{=}{a}^{1/{\alpha}} {V}_t^{(\alpha)},\qquad  a>0,
\end{displaymath}
where the constant $a$ is dimensionless. Then the corresponding property for the new
process $\overline{V}^{(\alpha)}_t$ takes the form
\begin{displaymath}
\overline{V}^{(\alpha)}_{at}\stackrel{d}{=}a \overline{V}_t^{(\alpha)},\qquad  a>0.
\end{displaymath}
This once again confirms their different behavior.

Now consider the subordinated process
$X(\overline{V}^{(\alpha)}_t)$ in which for simplicity a random
process $X(t)$ is ordinary Brownian, and the directing process is
$\overline{V}^{(\alpha)}_t$.  Let the subordinated process
$X(\overline{V}^{(\alpha)}_t)$ with such a subordinator have a
probability density $\hat{p}_\alpha(x,t)$. The relationship
between the probability densities of parent and directing
processes is expressed in the integral form
\begin{equation}
\hat{p}_\alpha(x,t)=\int^\infty_0
p_1(x,t\xi)\,dg_\alpha(\xi)\,,\label{eq19a}
\end{equation}
where the probability distribution $g_\alpha(\xi)$ is described by
the Laplace transform
\begin{displaymath}
G_\alpha(s)=\int^\infty_{0}\exp\{-s\xi\}\,dg_\alpha(\xi)=\exp\{-(As)^\alpha\}
\end{displaymath}
with $s\geq 0$ and $A>0$. The probability density $p_1(x,t)$ describes the normal
diffusion. The anomalous diffusion $X(\overline{V}^{(\alpha)}_t)$ gives the stretched
exponential function of relaxation
\begin{equation}
\left\langle
e^{-k\overline{V}^{(\alpha)}_t}\right\rangle=e^{-c_\alpha t^\alpha
k^\alpha},\qquad 0<\alpha<1, \quad c_\alpha>0.\label{eq19b}
\end{equation}
In the next section we will demonstrate how the combined laws of
relaxation can appear in the evolution of relaxing physical
systems due to a further development of the subordination
approach.

\section{Subordination by two random processes}\label{par3}
Let two independent sequences $\{T^{(1)}_i\}$ and $\{T^{(2)}_i\}$,
$i=1,2,\dots$ consist of nonnegative, independent and identically
distributed random variables. The random variables $T^{(1)}_i$ and
$T^{(2)}_i$ are attracted to $\alpha$-stable laws with different
indices. Let the temporal variables $T^{(1)}(n)$ and $T^{(2)}(n)$
be a sum of the sequences of time intervals, $T^{(1)}_i$ and
$T^{(2)}_i$ respectively. The counting process $N_t$  describes
the number of the particle jumps performed up to time $t$. The
position of the particle after $n$ jumps is a sum of the jumps
$R_i$. The total distance reached by a particle during time $t$ is
defined by the number of the jumps by means of the counting
process $N_t$. In this case the process $N_t$ is something like
the operational time. Since the sequences of the time intervals
are independent on each other, their subordinators will be such
too. To pass to the continuous limit, the probability density of
the obtained subordinate process should be expressed as
\begin{eqnarray}
&&p_{\alpha,\,\beta}(x,t)=\nonumber\\
&&=\int_0^\infty\int_0^\infty
f_\alpha(\tau_2,t)\,g\Bigl(x,\frac{\tau_1+\tau_2}{2}\Bigr)\,
f_\beta(\tau_1,t)\,d\tau_1\,d\tau_2\,,\label{21}
\end{eqnarray}
where the variables $\tau_1$ and $\tau_2$ correspond to the two-time scale subordination
for the parent process with the probability density $g(x,\tau)$. Here, we also have an
anomalous diffusion.

In the case of the nonbiased random walk the process $X(\tau)$ belonging to the class of
$\gamma$-stable L$\acute{\rm e}$vy process has the following characteristic function
\begin{displaymath}
\left\langle e^{ikX(\tau)}\right\rangle=e^{-c_\gamma k^\gamma\tau},\quad c_\gamma>0.
\end{displaymath}
If the parent process is directed by the two subordinators, then the relaxation function
takes the form
\begin{eqnarray}
&&\phi_{\rm two}(t)=\nonumber\\ &&=\int^\infty_0\int^\infty_0
e^{-c_\gamma
k^\gamma(\tau_1+\tau_2)/2}\,f_\alpha(\tau_1,t)\,f_\beta(\tau_2,t)\,d\tau_1\,d\tau_2.
\label{22}
\end{eqnarray}
As a result, we obtain a combined process.

Now, we look at the trapping reaction with the randomly moving traps. According to the
experimental data \cite{D}, one can prepare the corresponding experimental system so that
after activating the traps, the reactant concentration will relax in correspondence with
a combined law. This law may be a product of some (in the simplest case, two) well-known
relaxation laws (such as exponential, stretched exponential and so on). From the above
consideration it follows that the relaxation function for the trapping reaction can be
expressed in terms of
\begin{eqnarray}
\phi_{\rm TR}(t)&=&\left\langle
e^{-kV^{(\alpha)}_t-k\overline{V}^{(\beta)}_t}\right\rangle=
\left\langle e^{-kV^{(\alpha)}_t}\right\rangle\left\langle
e^{-k\overline{V}^{(\beta)}_t} \right\rangle=\nonumber \\
&=&e^{-c_\alpha k^\alpha t}\cdot e^{-c_\beta t^\beta
k^\beta},\,0<\alpha,\beta<1,\quad c_\alpha,c_\beta>0.\label{23}
\end{eqnarray}
To sum up, the relaxation function is written as a simple product
of the ordinary exponential and stretched exponential functions.
Two random processes are present at the reaction: 1) reactant
walks in a medium; 2) the traps appear randomly. If these
processes are independent, then their contribution in the
relaxation function can be a product of two relaxation
dependencies. One of them gives a simple exponent, and another
tends to a stretched exponential function. The indices $\alpha$
and $\beta$ permit one to distinguish the random processes, but we
do not define concretely what component (reactant or traps) leads
to, for example, the stretched-exponent contribution because this
depends on the experimental situation. We describe the most
general picture of this phenomena from the probabilistic formalism
of limit theorems.

Why is subordination important here? Because the reactant walks is
(sub)diffu-sive, and (sub)diffusion is a result of subordination
of one random process by another \cite{PirSW}. In like manner this
relates to traps. The nontriviality of our analysis is due to the
fact that the effective stochastic process in time at a fixed
point in space becomes non-Markovian. For a non-Markovian process,
calculation of any history dependent quantity such as functionals
of trajectories is extremely hard barring a few special  cases
(see more details, for example, in \cite{PirSW}). Over past decade
the probabilistic formalism of limit theorems has made a very
great advance on the analysis of non-Markovian processes. We just
apply it for the treatment of Djordjevi$\check{\rm c}$'s
experiment.

It should be emphasized that the subordination approach answers an
important question: what a stochastic mechanism stands behind the
combined law of relaxation in the studied system. We believe that
this is caused by a multi-time scale subordination of random
physical processes in such systems. Moreover, this is not the only
physical development of multi-time scale subordination. In the
next we consider how the subordination can influence on a state
equation of continuous media.

\section{State equation for continuous media with memory}\label{par4}
Any continuous medium at any point ${\bf x}=(x_1,x_2,x_3)\in {\bf R}^3$ and any time $t$
is determined by the velocity vector ${\bf v}({\bf x},t)=(v_1,v_2,v_3)$ , the pressure
$p({\bf x},t)$ and the density $\rho({\bf x},t)$. The characteristic quantities are
connected with each others by means of the Navier-Stokes equation
\begin{displaymath}
\rho\frac{d{\bf v}}{dt}+\nabla p=\mu\Delta {\bf v},
\end{displaymath}
the continuum equation
\begin{displaymath}
\frac{\partial\rho}{\partial t}+{\rm div}\,\rho {\bf v}=0
\end{displaymath}
and the state relation
\begin{displaymath}
F(p,\rho,{\bf v},{\bf x},t)=0,
\end{displaymath}
where $\nabla=\left(\frac{\partial}{\partial x_1},\frac{\partial}{\partial
x_2},\frac{\partial}{\partial x_3}\right)$ is the Hamiltonian operator, $\mu$ the
viscosity, $\Delta$ the Laplace operator, and $F$ is an operator (see \cite{LL}).

Following \cite{VRS}, for a start we assume that the relaxation process in a continuous
medium obeys a simple exponential law
\begin{equation}
dh/dt=-(h-h_0)/\theta\qquad\mathrm{or}\qquad
h=h_0+(h(0)-h_0)e^{-t/\theta},\label{eq31}
\end{equation}
where we do not define concretely the nature of relaxation, and
$\theta$ is the relaxation time of the parameter $h$
characterizing this medium. Next, a wave (sound or something like)
is propagated in the medium. If the wave period is $T\gg\theta$,
the media has time to accommodate oneself to wave changes, and the
wave will advance with the velocity $c_0^2=(\partial
p/\partial\rho)_{h_0}$ (here $h_0$ corresponds to a value of $h$
in this wave). If on the contrary $T\ll\theta$, then the parameter
$h$ will be ``frozen'' to its value $h_{00}$ without any wave, and
the wave velocity is $c_\infty^2=(\partial
p/\partial\rho)_{h_{00}}$. Suppose that the wave results in small
perturbations of the density $\rho=\rho_0+\rho\,'$ and the
pressure $p=p_0+p\,'$ for the given medium.

Let the state relation take the form $p=p(\rho,h)$. Expand it in a
series about small perturbations:
\begin{equation}
p=p(\rho,h)\approx p_0+\Bigl(\frac{\partial
p}{\partial\rho}\Bigr)_{h_0}(\rho-\rho_0)+\Bigl(\frac{\partial
p}{\partial h}\Bigr)_{\rho_0}(h-h_0).\label{eq32}
\end{equation}
As the equilibrium state has $h_0=g_0(\rho)$, therefore
\begin{equation}
p(\rho,h_0)\approx p_0+\Bigl(\frac{\partial
p}{\partial\rho}\Bigr)_{h_{00}}(\rho-\rho_0)+\Bigl(\frac{\partial
p}{\partial h}\Bigr)_{\rho_0}(h_0-h_{00}).\label{eq33}
\end{equation}
It follows from this that
\begin{equation}
\Bigl(\frac{\partial
p}{\partial\rho}\Bigr)_{h_0}=\Bigl(\frac{\partial
p}{\partial\rho}\Bigr)_{h_{00}}+\Bigl(\frac{\partial p}{\partial
h}\Bigr)_{\rho_0}\, \frac{\partial
h_0}{\partial\rho}\,.\label{eq34}
\end{equation}
Next, we substitute the expression (\ref{eq34}) to the expansion (\ref{eq32}) and
differentiate the obtained result with respect to $t$. Using the equation (\ref{eq31})
one can find
\begin{equation}
\frac{dp\,'}{dt}=c^2_\infty\frac{d\rho\,'}{dt}-\Bigl(\frac{\partial
p}{\partial g}\Bigr)_{\rho_0}\,\frac{h-h_0}{\theta}\,.\label{eq35}
\end{equation}
The expression (\ref{eq35}) shows a connection between increments
of $p\,'$ and $\rho\,'$ in the wave and the deviation of the
parameter $h-h_0$ from its equilibrium value. In order to get an
equation depending only on $p\,'$ and $\rho\,'$, we divide
Eq.(\ref{eq32}) on $\tau$ and sum the result with Eq.(\ref{eq35}).
Thus we arrive at the sought-for equation:
\begin{equation}
\frac{dp\,'}{dt}+\frac{p\,'}{\theta}=c^2_\infty\frac{d\rho\,'}{dt}+c_0^2\frac{\rho\,'}
{\theta}\,.\label{eq36}
\end{equation}
It is not difficult to establish that this equation is equivalent to the integral
relation
\begin{equation}
p\,'=c^2_0\rho\,'+(c^2_\infty-c^2_0)\int^t_{-\infty}\exp\left(-\frac{t-t_1}{\theta}\right)
\,\frac{d\rho\,'}{dt_1}\,dt_1\,.\label{eq37}
\end{equation}
From Eq.(\ref{eq37}) it is seen that the medium has memory effects.

For the derivation of Eq.(\ref{eq37}) we accepted the relaxation equation of type
(\ref{eq31}). As a result, the kernel of Eq.(\ref{eq37}) has the exponential form.
Generally, the relaxation equation can be more complex (for example, for polymers it is
non-exponential), and therefore the corresponding kernel may be other. In particular, for
fractional systems it is written as a power function, and the corresponding state
equation takes the fractional form of differential equations. Consider the feature in
more details below.

A broad class of continuous media with a long-term memory (for example, polymer fluids,
viscoelastic materials, etc. \cite{BT,N}) is described more adequately by the state
equation in the fractional form of type
\begin{equation}
\frac{\partial^\alpha p_\alpha}{\partial t^\alpha}
+\frac{p_\alpha}{\theta_1}=c^2_\infty\frac{\partial^\beta\rho_\beta}{\partial
t^\beta}+c_0^2\frac{\rho_\beta}{\theta_1}, \label{eq321}
\end{equation}
where $\partial^\alpha/\partial t^\alpha$ is the fractional
derivative in the sense of Caputo \cite{IP}. Here $c_0$ is the
wave velocity, when the wave period is more than relaxation time
$\theta_1$, and $c_\infty$ is the wave velocity in the opposite
case. To obtain Eq.(\ref{eq321}) from Eq.(\ref{eq36}), we present
the pressure and the density as a result of subordination, namely
\begin{eqnarray}
p_\alpha({\bf x},t)&=&\int^\infty_0f_\alpha(\tau_1, t)p({\bf
x},\tau_1)d\tau_1,\label{eq331a}\\ \rho_\beta({\bf
x},t)&=&\int^\infty_0f_\beta(\tau_2, t)\rho({\bf
x},\tau_2)d\tau_2,\label{eq331b}
\end{eqnarray}
where $f_\alpha(\tau_1,t)$ (or $f_\beta(\tau_2,t)$) is the probability density of the
inverse-time  $\alpha$ (or $\beta$)-stable L$\acute{\rm e}$vy subordinator with
$0<\alpha,\beta<1$. In other words, both the pressure and the density of such a medium
are subordinated in different ways. From the physical point of view the subordination
accounts that there are temporal (random) intervals, when the pressure and the density
does not change. In any other case the random jumps in density and pressure occur. In
particular, if $\alpha,\beta=1$, then the values change continuously.

According to \cite{S2}, the Laplace transform in time, for the pressure and the density
in the medium with the long-term memory, gives
\begin{equation}
\bar{p}_\alpha({\bf x},s)=s^{\alpha-1}\bar{p}({\bf
x},s^\alpha),\quad \bar{\rho}_\beta({\bf
x},s)=s^{\beta-1}\bar{\rho}({\bf x},s^\beta).\label{eq341}
\end{equation}
Hence, Eq.(\ref{eq36}) in the Laplace space reads
\begin{eqnarray}
s\bar{p}({\bf x},s)&-&p({\bf x},0)+\frac{\bar{p}({\bf x},s)}{\theta_1}=\nonumber\\
&=&c^2_\infty s\bar{\rho}({\bf x},s)-c^2_\infty\rho({\bf
x},0)+c_0^2\frac{\bar{\rho}({\bf x},s)}{\theta_1}.\label{351}
\end{eqnarray}
The subordination in different ways means that we should accept
\begin{eqnarray}
\int_0^\infty\left(\frac{\partial p({\bf x},\tau_1)}{\partial\tau_1} +\frac{p({\bf
x},\tau_1)}{\theta_1}\right)f_\alpha(\tau_1,t)\,d\tau_1=\nonumber\\
=\int^\infty_0\left(c^2_\infty\frac{\partial\rho({\bf
x},\tau_2)}{\partial\tau_2}+c_0^2\frac{\rho({\bf
x},\tau_2)}{\theta_1}\right)\,
f_\beta(\tau_2,t)\,d\tau_2.\label{361}
\end{eqnarray}
Then the Laplace transform of the latter expression takes the form
\begin{eqnarray}
s^{\alpha-1}(s^\alpha\bar{p}({\bf x},s^\alpha)-p({\bf
x},0))+\frac{\bar{p}_\alpha({\bf x},s)}{\theta_1}=\nonumber\\
c^2_\infty s^{\beta-1}(s^\beta\bar{\rho}({\bf
x},s^\beta)-\rho({\bf x},0))+c_0^2\frac{\bar{\rho}_\beta({\bf
x},s)}{\theta_1}.\label{371}
\end{eqnarray}
After simple algebraic transformations we have
\begin{eqnarray}
s^\alpha\bar{p}_\alpha({\bf x},s)-s^{\alpha-1}p({\bf
x},0)+\frac{\bar{p}_\alpha({\bf x},s)}{\theta_1}=\nonumber\\
c^2_\infty s^\beta\bar{\rho}_\beta({\bf x},s)-c^2_\infty
s^{\beta-1}\rho({\bf x},0)+c_0^2\frac{\bar{\rho}_\beta({\bf
x},s)}{\theta_1}.\label{381}
\end{eqnarray}
and the Laplace inverse of this expression leads directly to
Eq.(\ref{eq321}). The subordination of type (\ref{eq331a}) and
(\ref{eq331b}) does not change the form of the Navier-Stokes
equation, but in the continuum equation the time derivative for
the density becomes fractional.

Finally, we consider the situation where both the Navier-Stokes equation and the
continuum equation are one-dimensional, namely
\begin{eqnarray}
\rho_0\frac{\partial v}{\partial t}&+&\frac{\partial
p_\alpha}{\partial x}=\mu\frac{\partial^2 v}{\partial x^2},\quad
v\equiv v_1,\quad x\equiv x_1,\label{eq391a}\\ &&\nonumber\\
\frac{\partial^\beta\rho_\beta}{\partial
t^\beta}&+&\rho_0\frac{\partial v}{\partial x}=0,\quad \rho\equiv
{\rm const}>0.\label{eq391b}
\end{eqnarray}
For $\alpha=\beta$ and $c_0<<1$ Eq.(\ref{eq321}) tends to the following expression
\begin{equation}
\theta_1\frac{\partial^\alpha p_\alpha}{\partial t^\alpha}
+p_\alpha=\theta_1
c^2_\infty\frac{\partial^\alpha\rho_\alpha}{\partial
t^\alpha}.\label{eq362}
\end{equation}
Let the viscosity $\mu$ be equal to zero. From Eqs.(\ref{eq391a})
and (\ref{eq391b}) we have
\begin{displaymath}
\frac{\partial}{\partial t}\frac{\partial^{\alpha}\rho_\alpha}{\partial
t^{\alpha}}=\frac{\partial^2 p_\alpha}{\partial x^2}.
\end{displaymath}
Using the state equation (\ref{eq362}), the equation for the
pressure takes the form
\begin{equation}
\frac{\partial p_\alpha}{\partial
t}+\theta_1\frac{\partial}{\partial
t}\frac{\partial^{\alpha}p_\alpha}{\partial t^{\alpha}}=\theta_1
c^2_\infty\frac{\partial^2 p_\alpha}{\partial x^2}.\label{eq372}
\end{equation}
The presented case is interesting because it leads to the
fractional wave-diffusion equation (\ref{eq372}), although the
Navier-Stokes equation is not fractional.  The fractional
derivative appears in Eq.(\ref{eq372}), due to the fractional
state equation and the fractional continuum equation.

\section{Conclusions}\label{par5}
We have shown how the combined law of relaxation can be derived
from the diffusion model based on the CTRW consideration and
illustrated the role of the two-time scale subordination. Our
analysis demonstrates that the type of relaxation function depends
entirely on the subordinator. The subordibators may be different
and enough complicated. If the operational time is an inverse-time
L$\acute{\rm e}$vy $\alpha$-stable subordinator, the relaxation
function becomes Mittag-Leffler's. The operational time of type
$\overline{V}^{(\alpha)}_t$ ``transforms'' the relaxation function
into the stretched exponential form. The multi-scale subordinator
allows one to combine some laws of relaxation as their product.
The result supports the conclusion that the subordination as a
transformation from the physical time $t$ to the operational time
$\tau$ is responsible for the anomalous properties of complex
systems. Using this approach to a similar problem in continuous
mechanics, we notice a special role of the state relation. Due to
the multi-scale subordination the relation takes the fractional
form. Consequently, the continuum equation becomes mathematically
more complicated than one in the ordinary mechanics.

\section*{Acknowledgments}
AS is grateful to the Institute of Physics and the Hugo Steinhaus Center for Stochastic
Methods for pleasant hospitality during his visit in Wroc{\l}aw University of Technology.

\end{document}